\documentclass[aps,pra,twocolumn,showpacs,groupedaddress,nofootinbib]{revtex4}
\usepackage{bbm,amsmath,amssymb,amsbsy,graphicx,subfigure,times}

\usepackage{hyperref}
\usepackage[latin1]{inputenc}
\newcommand{\gr}[1]{\boldsymbol{#1}}
\newcommand{\ket}[1]{|#1\rangle}
\newcommand{\bra}[1]{\langle#1|}

\newcommand{\eq}[1]{Eq.~(\ref{#1})}

\newcommand{\sig}{{\gr\sigma}}

\begin{document}
\title{Experimentally friendly bounds on non-Gaussian entanglement from second moments}

\date{January 15, 2009}
\author{Gerardo Adesso}
\affiliation{School of Mathematical Sciences, University of Nottingham,
University Park,  Nottingham NG7 2RD, UK.}
\affiliation{Dipartimento di Matematica e Informatica, Universit\'a
degli Studi di Salerno, Via Ponte Don Melillo, 84084 Fisciano (SA),
Italy}


\begin{abstract}
We demonstrate that the entanglement in a class of two-mode
non-Gaussian states obtained by subtracting photons from Gaussian
twin beams can be bounded from above and from below by functionals
of the second moments only. Knowledge of the covariance matrix thus
suffices for an entanglement quantification with appreciable precision. The absolute error in the entanglement estimation scales with the non-Gaussianity of the considered states.
\end{abstract} \maketitle


\section{Introduction}

Interfaces between light and matter are key building blocks of a
future {\em quantum web}, a global secure communication network
where the manipulation and transmission of information are regulated
by the quantum laws \cite{qinternet}. The transfer of quantum states
and the distribution of correlations across the interfaces are
enabled by the common mathematical language of canonically conjugate
observables with continuous spectrum, such as the quadratures of
light and the collective spin components of atomic ensembles. It is
thus very fascinating to witness how the second generation of
quantum information research is focusing more and more, from both
theoretical and experimental viewpoints, on the characterization of
continuous variable (CV) entanglement and its applications for
communication, computation and metrology \cite{brareview}.

Out of the infinite-dimensional Hilbert space of CV states, a
special class of states has played a prominent role in recent years:
Gaussian states.
Their mathematical treatment is advantageous thanks to a compact
formalism based on symplectic analysis, and  a very accurate degree
of control is reached in their experimental realizations with light
and matter \cite{ourreview}. There are however many tasks which are
impossible by using only the Gaussian states and operations toolbox
(e.g., entanglement distillation \cite{nogo} and universal quantum
computation \cite{lloydmenicucci}), and many other tasks which can
be sharply improved by suitably resorting to some non-Gaussianity
(e.g. CV teleportation \cite{fabiotele} and loss estimation
\cite{ourloss}). These premises have spurred astonishing progresses
in the experimental engineering of non-Gaussian states, such as Fock
states and states obtained by deGaussifying Gaussian resources via
addition and/or subtraction of single photons \cite{belliniecc}. It
has been in particular verified experimentally that a photon
subtraction from a two-mode squeezed entangled Gaussian state leads
to an enhancement of the entanglement (at fixed squeezing)
\cite{ourj}, and it is known that such resource might be used for a
more performant quantum teleportation of classical and nonclassical
states (a primitive of the quantum internet)
\cite{photsub,fabiotele,kitagawa} and for yet-to-be-achieved
demonstrations of loophole-free Bell tests of nonlocality
\cite{homobell}. The bottleneck for unleashing the power of
non-Gaussian CV quantum technology has remained the quantitative
characterization of entanglement in states which deviate from
Gaussianity, a crucial step to evaluate and control their usefulness
for applications and the {\em bona-fide}ness of their preparation.
While for Gaussian states all the information is encoded in the
second moments of the canonical operators (collected in the
covariance matrix),  for any other state an infinite hierarchy of
moments is in principle needed for an exact entanglement
quantification \cite{schukin}. This translates, in experimental
terms, into the demand for a complete state tomography
\cite{homotomo}, a process which is time- and resource-consuming especially  for two or more modes.

In this paper we provide an advance in the characterization of
non-Gaussian entanglement which reduces the complexity of its
experimental determination exactly to the same level of Gaussian
states. This cannot be possible for {\em any generic} non-Gaussian
state: hence here we focus on the important class of
photon-subtracted states (PSS) under quite realistic conditions
\cite{photsub,kitagawa}, which represent the preferred resources for
most current and future applications requiring non-Gaussianity.
Combining results from the extremality of Gaussian states
\cite{extra} with an analysis of quadrature correlations, we derive
analytical lower and upper bounds which individuate the entropy of
entanglement of pure two-mode PSS
 (obtained from Gaussian twin beams by conditional subtraction
of $k$  photons per beam) within a very narrow band, with relative
error vanishing with increasing entanglement. The bandwidth is
linked to the degree of non-Gaussianity \cite{parisentro} of the
analyzed PSS. The results are extended to bound the entanglement of
formation of a class of mixed PSS obtained by means of realistic ``on/off''
type photodetectors. Crucially, all these bounds are only functions
of the second moments of the non-Gaussian states. A novel method for
the fast and reliable reconstruction of the complete covariance
matrix of optical two-mode CV states has been recently demonstrated
\cite{porzio,noteatom}: our result proves that it can be applied to real
deGaussified PSS as well and it is enough to quantitatively estimate
entanglement with surprisingly high accuracy.

\section{Preliminaries} We deal with a CV system of $N=2$
bosonic modes, associated to an infinite-dimensional Hilbert space
${\cal H}= F_1 \otimes F_2$ \cite{ourreview}. Here $F_i$ is the Fock
space of each individual mode $i$, described by the ladder operators
$\hat {a}_i ,\,\hat {a}_i ^\dag $ satisfying the canonical
commutation relations $[\hat {a}_i ,\,\hat {a}_j
^\dag]=\delta_{ij}$. We can collect the field quadrature operators
 into the vector $\hat{X} = \{\hat
q_1, \hat p_1, \hat q_2, \hat p_2\}$, with $\hat {q}_i =\hat {a}_i
+\,\hat {a}_i ^\dag$ and $\hat {p}_i =\hat {a}_i -\,\hat {a}_i
^\dag$. The exact description of a generic CV (non-Gaussian) state
requires arbitrary-order moments of the canonical operators. In
general, the first moments $\bar X\equiv (\langle\hat X_{1}
\rangle,\langle\hat X_{1}\rangle, \langle\hat X_{2}\rangle,
\langle\hat X_{2}\rangle)$ can be adjusted by local displacements
without affecting entanglement: they will be set to zero without
loss of generality. Two-mode Gaussian states are henceforth
completely specified by the $4 \times 4$ real symmetric covariance
matrix (CM) $\gr{\sigma}$ of the second moments $\sigma_{ij}=
\langle \hat{X}_i \hat{X}_j + \hat{X}_j \hat{X}_i \rangle /2 -
\langle \hat{X}_i \rangle \langle \hat{X}_j \rangle$
\cite{ourreview}. The CM elements for an arbitrary two-mode CV state
can be efficiently reconstructed in the lab by means of homodyne
detections \cite{francesi,porzio}.  We recall that any physical
two-mode CM $$\sig={{\gr \alpha_1\,\,\gr\gamma} \choose {\gr
\gamma^T\, \gr\alpha_2}}$$ can be converted by local unitaries into a
standard form \cite{duan} where $\gr\alpha_i={\rm
diag}\{a_i,\,a_i\}$ and $\gr\gamma={\rm
diag}\{\gamma_x,\,\gamma_p\}$ with $a_i \ge 1$, $\gamma_x \ge
|\gamma_p| \ge 0$.

Entanglement of a pure bipartite state $\ket{\psi}$ is universally
quantified by the Von Neumann entropy of the reduced density matrix
of each subsystem (entropy of entanglement $E$) \cite{plevir}. The
optimal convex-roof extension of $E$ for a mixed state $\varrho$
defines the entanglement of formation \cite{plevir}  $E_F(\varrho) =
\inf_{\{p_i,\psi_i\}} \sum_i p_i E(\psi_i)$, where the infimum runs
over all pure-state decompositions of  $\varrho = \sum_i p_i
\ket{\psi_i}\!\bra{\psi_i}$.  Another popular measure of
entanglement for pure and mixed states, whose computation is in
general simpler, is the logarithmic negativity \cite{plevir,vidwer}  $E_N =
\log\big[{\rm tr} |\varrho^{T_i}|\big]$ \cite{notelog}, where the
partial transposition $\varrho^{T_i}$ is obtained from $\varrho$ by
transposing the degrees of freedom of one subsystem only. For pure
states $E_N \ge E$. An important result in CV entanglement theory is
the `extremality' of Gaussian states: for any CV state $\varrho$
with second moments given by $\sig$, the corresponding Gaussian
state $\varrho^G$ defined by the same CM has smaller entanglement
(quantified by a continuous and strongly superadditive measure) than
$\varrho$ \cite{extra}. It is also known that for arbitrary two-mode
Gaussian states $E_F$ is computable, additive, and strongly
superadditive
  \cite{efsym,marian}.

\section{Covariance matrix and entanglement of pure photon
subtracted states}  The starting point for the definition of
``ideal'' PSS is a pure two-mode squeezed Gaussian state (or `twin
beam' in the optical language), $\ket{\psi_0 (r_0)} = \sum_n
\lambda^n \sqrt{1-\lambda^2} \ket{n,n}$, where $\lambda=\tanh r_0$
and the positive $r_0$ is the squeezing degree. The CM $\sig_0$ for
this state is already in standard form with $a_1=a_2=\cosh 2
r_0,\,\gamma_x=-\gamma_p=\sinh 2 r_0$. The beam $1$ ($2$) of
$\ket{\psi_0}$ is let to interfere, via a beam splitter with
transmittivity $T$ (preferably $T$ close to unity), with a vacuum
mode $1^\prime$ ($2^\prime$). The output is a four-mode Gaussian
state of modes $1,1^\prime,2,2^\prime$. A photon-number-resolving
detection of exactly $k$ photons in each of the two beams $1^\prime$
and $2^\prime$, conditionally projects the state of modes $1,2$ into
a pure symmetric non-Gaussian state \cite{notesym}, given in the
Fock basis by $\ket{\psi_k} = \sum_{n=k}^{\infty} c_n^{(k)}
\ket{n-k,n-k}$, with \cite{kitagawa}
\begin{equation}\label{eq:pcn}
c_n^{(k)} = (T \lambda )^{n-k} \left(
\begin{array}{c}
 n \\
 k
\end{array}
\right) \sqrt{\frac{\left(1-T^2 \lambda
   ^2\right)^{2 k+1}}{\,
   _2F_1\left(-k,-k;1;T^2 \lambda
   ^2\right)}}\,,
   \end{equation}
   where $_2F_1$ denotes the Gauss hypergeometric function.
We define an effective squeezing parameter $r$ such that $$T \lambda
\equiv z =\tanh r.$$ Thus $\ket{\psi_k} = {\cal N}_k^{-1/2}
\hat{a}_1^k \hat{a}_2^k \ket{\psi_0 (r)}$, where \[\begin{split}
{\cal N}_k &=
\sum_{n=k}^{\infty} [c_n^{(0)} {n \choose k} k!]^2  \\ &= k!^2\,
_2F_1\left(k+1,k+1;1;\tanh ^2 r\right)
   \cosh^{\!-2}\! r \tanh^{2 k}\! r\,.\end{split}\]
   The logarithmic negativity of
   the PSS states for arbitrary $k$ can be computed in simple closed
   form via the formula \cite{kitagawa} $E_N(\psi_k) = 2 \log [\sum_{n=k}^{\infty}
   c_n^{(k)}]$ and reads in our notation
   \begin{equation}\label{eq:pslogneg}
   E_N(\psi_k) = - \log[ (1-z)^{2 k+2} \,
   _2F_1\left(k+1,k+1;1;z^2\right) ]\,.
   \end{equation}
Notice that $E_N(\psi_k)$ increases with $k$: iteration of the
photon subtraction process further enhances the entanglement
compared to the original Gaussian instance. We aim at an estimate of
the entropy of entanglement of PSS for any $k$, and specifically at
accurate bounds on this universal entanglement quantifier which can
be measured experimentally with high efficiency. We will now derive
the CM of PSS states in closed form and show that it contains enough
information, straightforwardly accessible in the lab, for the
desired task.

\begin{figure*}[bt]
 {\includegraphics[width=7.5cm]{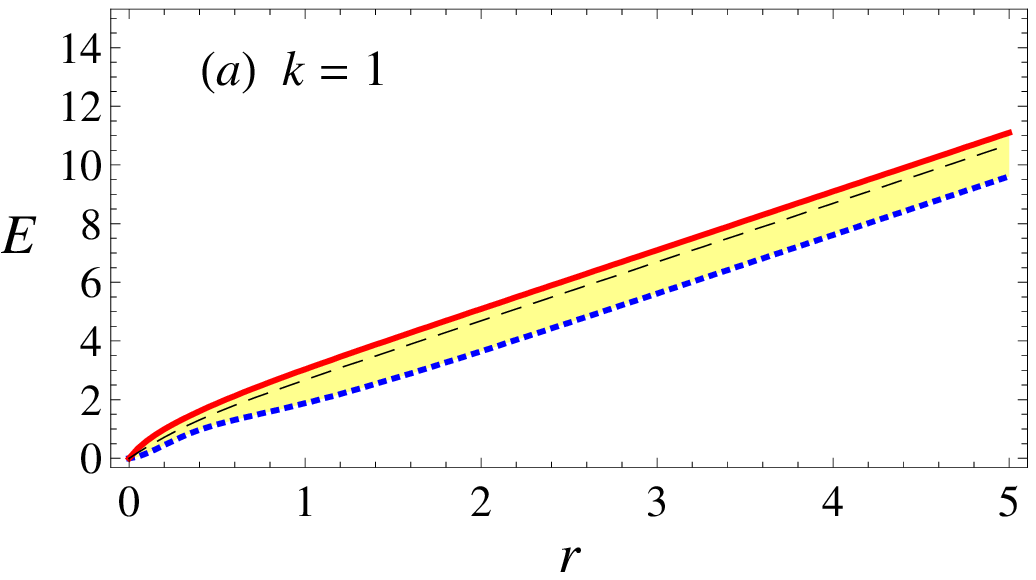}}\hspace*{.5cm}
 {\includegraphics[width=7.5cm]{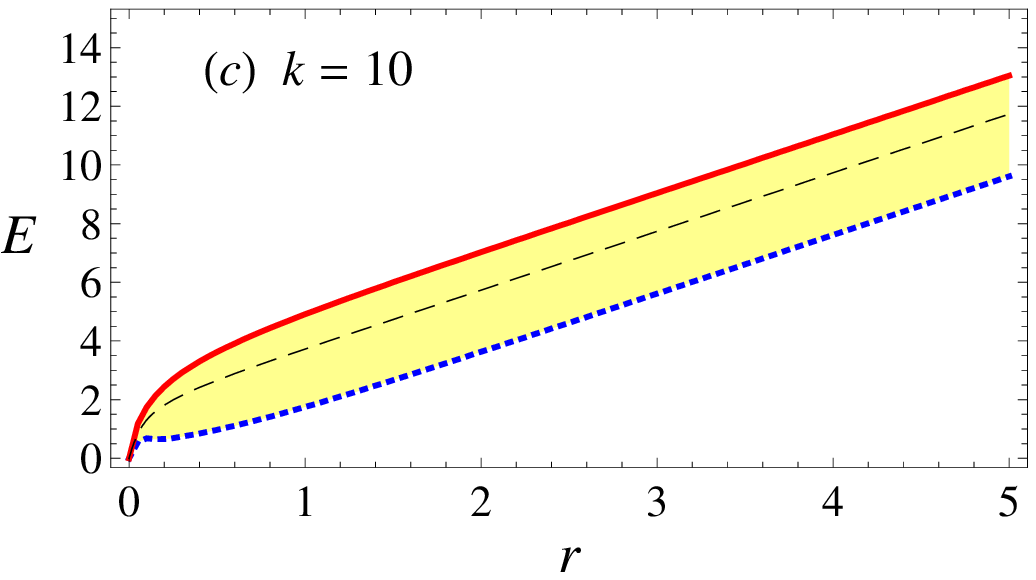}}\\\vspace*{.25cm}
 {\includegraphics[width=7.5cm]{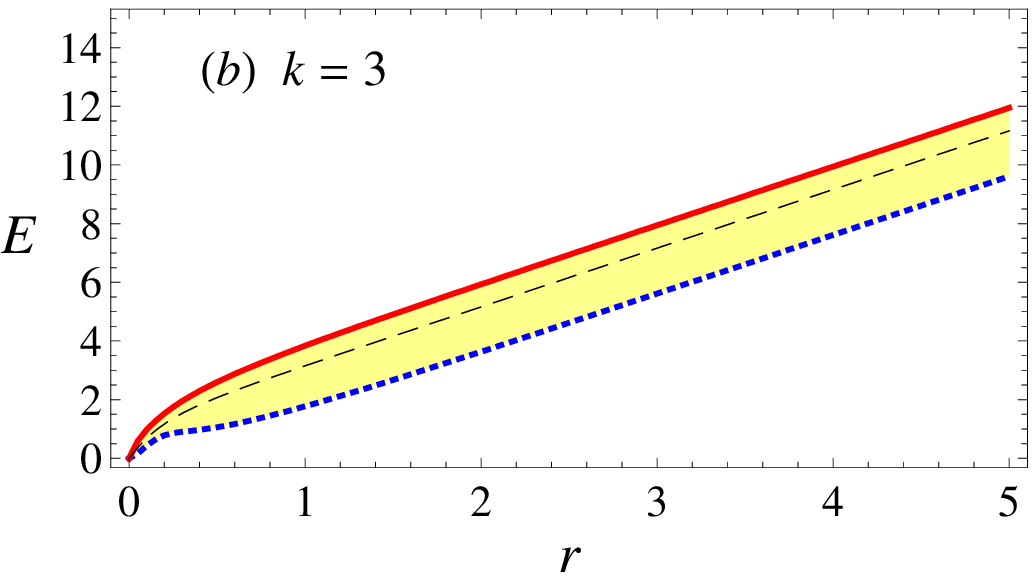}}\hspace*{.5cm}
 {\includegraphics[width=7.5cm]{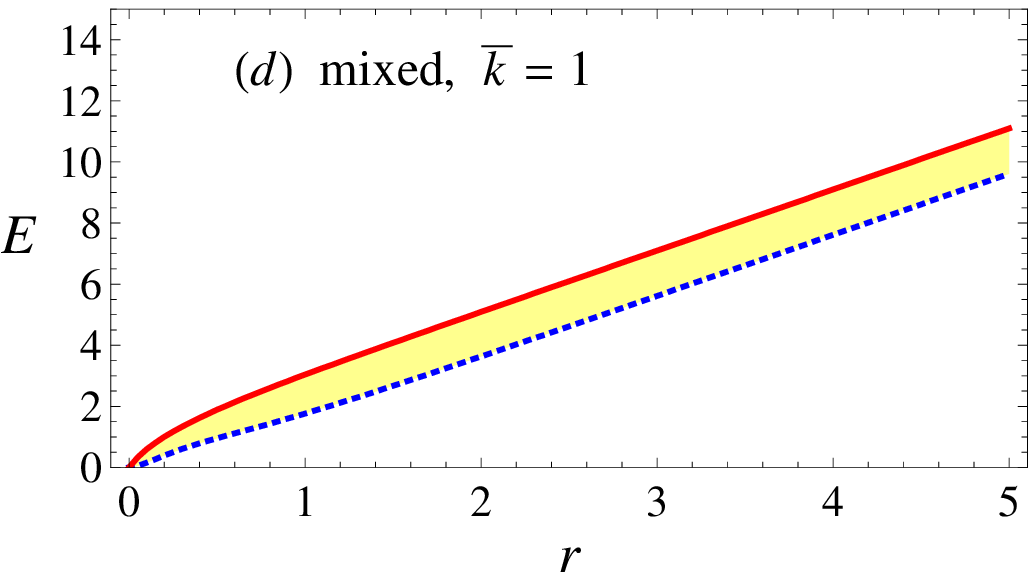}}
\caption{(color online). Upper (solid) and lower (dotted) bounds on
the entanglement  of PSS obtained from second moments, plotted as
functions of the squeezing $r$. The entanglement of formation
is
individuated within the narrow shaded region between the bounds.
Panels [{\it(a)--(c)}] depict ideal PSS ${\ket{\psi_k}}$, whose
logarithmic negativity is shown as well for reference (thin dashed
line). Panel {\it (d)} depicts a more realistic  PSS $\varrho_{{\bar k} =
1}$ modeled as a binomial mixture of the Gaussian twin beam and the
first four pure PSS, $\varrho_{{\bar k} = 1} = \sum_{k=0}^{n}
(1-p)^{n-k}p^k {n \choose k} \ket{\psi_k}\!\bra{\psi_k}$, with $n =
p^{-1}=4$; notice how only slight  modifications occur to the bounds
compared to the corresponding ideal case {\it (a)}.}
\label{fig:bounds}
\end{figure*}

Denoting by $\hat O$ an observable involving normally ordered
combinations of ladder operators on modes $1$ and $2$, we observe
that $\bra{\psi_k} \hat O \ket{\psi_k} = {\cal N}_k^{-1}
\bra{\psi_0} {\hat{a}_1^{\dag k}} {\hat{a}_2^{\dag k}} \hat O
\hat{a}_1^k \hat{a}_2^k\ket{\psi_0}$, that is, we can evaluate
expectation values of relevant operators in terms of normally
ordered higher moments computed on the unperturbed Gaussian twin
beams. Recalling that, for  $\ket{\psi_0(r)}$, the normally ordered
characteristic function reads \cite{brareview}
$\chi_0^{(N)}(\alpha_1,\alpha_1^\ast,\alpha_2,\alpha_2^\ast) =
\exp\{\sinh r [(\alpha_1\alpha_2+\alpha_1^\ast \alpha_2^\ast) \cosh
r - (|\alpha_1|^2+|\alpha_2|^2) \sinh r]\}$, we can write the
(normally ordered) moment generating formula:
\begin{eqnarray}\label{eq:momgen}
M_{il}^{jm}\!\!&=\!\!&\,_{\psi_{0\!}}\langle (\hat{a}_1^\dag)^j
(\hat{a}_1)^i (\hat{a}_2^\dag)^m (\hat{a}_2)^l \rangle_{\psi_0}
\\&=\!\!& \left.\frac{(-1)^{j+m}\,\,\partial^{i+j+l+m}}{\partial {\alpha_1}^i
\partial {\alpha_1^\ast}^j \partial {\alpha_2}^l \partial
{\alpha_2^\ast}^m}
\chi_0^{(N)}\!(\alpha_1,\alpha_1^\ast,\alpha_2,\alpha_2^\ast)\right|_{\begin{array}{l}
_{\!\alpha_{1,2}=0}\\ ^{\!\alpha_{1,2}^\ast=0}
\end{array}}\!.
\nonumber
\end{eqnarray}
The formula \eq{eq:momgen} can be readily applied to compute the
second moments of the canonical operators on our non-Gaussian
states. After some algebra, the CM $\sig_k$ of PSS of the form
$\ket{\psi_k}$ turns out to be that of a symmetric two-mode squeezed
thermal state, automatically in standard form, with $a_1=a_2\equiv
a^{(k)}={\cal N}_k^{-1} (M_{k+2,k}^{k,k} + M^{k+2,k}_{k,k} + 2
M_{k+1,k}^{k+1,k} + M_{k,k}^{k,k})$, and $\gamma_x = -\gamma_p
\equiv \gamma^{(k)} = {\cal
N}_k^{-1}(M_{k+1,k+1}^{k,k}+M^{k+1,k+1}_{k,k}+M_{k+1,k}^{k,k+1}+M_{k,k+1}^{k+1,k})$.
Explicitly:
\begin{eqnarray}
\label{eq:cx1} &\!\!\!\!\!\!\!a^{(k)}=& (\cosh r)^{2+4k}
[_2F_1(k+1,k+1;1;\tanh^2
r)]^{-1} \nonumber \\
&\!\!\!\!\!\!\qquad\times&[2(k+1)^2 \sinh^2 r \, _2F_1(-k,-k;2;\tanh^2 r) \nonumber \\
&\!\!\!\!\!\!\!& +\ _2F_1(-k,-k;1;\tanh^2 r)]\,; \\
\label{eq:cx3} &\!\!\!\!\!\!\!\gamma^{(k)}=& \frac{2  \tanh r (k+1)
\, _2F_1(k+1,k+2;1;\tanh^2 r)}{\,
   _2F_1(k+1,k+1;1;\tanh^2 r)}\,.
   \end{eqnarray}
We will now show how to extract from the CM elements a lower and an
upper bound on $E(\psi_k)$.

 {\noindent \em Lower
bound}.--- The extremality of Gaussian states \cite{extra} entails that \cite{noteconj}
\begin{equation}\label{eq:elow}
E^{\rm low}_k\equiv E_F(\varrho^G_k) \le E(\psi_k)\,,\end{equation}
where $\varrho^G_k$ is the mixed Gaussian state with CM $\sig_k$.
Explicitly, $E^{\rm low}_k = g[a^{(k)}-\gamma^{(k)}]$ where
\cite{efsym} $g(x)=\frac{(1+x)^2}{4x}\log
\left[\frac{(1+x)^2}{4x}\right]- \frac{(1-x)^2}{4x}\log
\left[\frac{(1-x)^2}{4x}\right]$.

{\noindent \em Upper bound}.--- The logarithmic negativity   of pure
PSS is already an upper bound for $E$, however it is defined in terms of all the moments of the state, and its experimental
determination (in real conditions requiring a complete state tomography) becomes rather demanding even when the states take very special forms \cite{ourj}. The
strength of our investigation is to derive a slightly looser upper
bound but which is a function of the second moments only of the PSS,
i.e. of $\sig_k$. We simply observe that for $k=0$, namely for
Gaussian twin beams, $E_N(\psi_0) = {\mathrm{arcsinh}}
[\gamma^{(0)}]$. The standard-form parameter $\gamma^{(0)}$
quantifies the maximum quadrature correlation between the two modes
\cite{carles}. It is tempting to postulate that, for PSS
$\ket{\psi_k}$ with an arbitrary degree $k$ of deGaussification, a
function of $\gamma^{(k)}$ may yield an overestimate of the actual
entanglement (which, we remark, is nontrivially encoded in
higher-order correlations  too). We can turn this
blurry bit of intuition into the following \\
{\noindent \em Theorem 1.} For all $k \ge 0$,
\begin{equation}\label{eq:eup}
E^{\rm up}_k \equiv \log[1+2 \gamma^{(k)}] \ge E_N(\psi_k) \ge
E(\psi_k)\,.\end{equation}

\noindent{\em Proof.} The rightmost inequality holds by definition.
Here we sketch the (quite technical) proof of the leftmost one, which
is one of the main results of this paper. The simple cases $k=0,\ 1$
can be proven by inspection, hence we specify here to arbitrary $k
\ge 2$. Using \eq{eq:pslogneg} and \eq{eq:cx3}, and exponentiating
both sides of the inequality, the problem reduces to proving that
$F^{(k)}(z) \equiv {_2F_1}\left(k+1,k+1;1;z^2\right)+4z(k+1) \,
{_2F_1}\left(k+2,k+1;1;z^2\right) - (1-z)^{-2 k-2}\ge 0$ where we
recall that  $\tanh r \equiv z  \in (0,\ 1)$. We can write
$F^{(k)}(z)$ as a power series,
$F^{(k)}(z)=\sum_{m=1}^{\infty}f_m^{(k)} z^m$, where
$f_m^{(k)}=\left[\left(1-(-1)^m\right) \left(2
   k+m+\frac{1}{2}\right)+1\right] {{\frac{1}{4}
   \left(-1+(-1)^m\right)+k+\frac{m}{2}} \choose k}^2
-{{2k+m+1} \choose {2k+1}}$. We observe that $f_{2j-1}^{(k)} \ge 0$
and $f_{2j}^{(k)} \le 0$ $\forall j \ge1$, and moreover
$f_{2j-1}^{(k)} \ge - f_{2j}^{(k)}$ $\forall j > k+1$. This entails
that by truncating the power series at $m=2k+2$ we discard a
positive remainder: $F^{(k)}(z) \ge {\tilde F}^{(k)}(z) \equiv
\sum_{m=1}^{2k+2}f_m^{(k)} z^m$. Let us now define a parametric
class of hypergeometric sums, $S_l^{(k)} = \sum_{m=1}^{2k+2}
{{2k+l-m+1} \choose {l}} f_m^{(k)}$. By means of Zeilberger's
algorithm \cite{zeilberger} one can verify that $S_l^{(k)} >0$
$\forall k\!\ge\!2,\ l\!\ge\!0$. We will now show that ${\tilde
F}^{(k)}(z) \ge S_0^{(k)} z^{2k+2} \ge 0$ to conclude the proof. We
take the ratio $R^{(k)}(z) = [{\tilde F}^{(k)}(z)]/[S_0^{(k)}
z^{2k+2}]$ and expand it in power series around $z= 1^{-}$:
$R^{(k)}(z) =1+\sum_{l=1}^{\infty}(-1)^l[S_l^{(k)}/S_0^{(k)}]
(z-1)^l$. But trivially $(z-1)^l = (-1)^l (1-z)^l$, and being $z \le
1$ the alternating sign is cancelled to yield $R^{(k)}(z) \ge 1$.
$\hfill \Box$

\begin{figure}[tb]
\includegraphics[width=8cm]{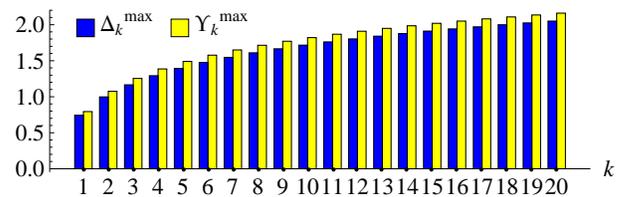}
\caption{(color online). Scaling as a function of $k$ of the
asymptotic absolute error $\Delta_k^{\max}$ on the estimate of
entanglement via second moments (dark bars) and of the asymptotic
entropic non-Gaussianity $\Upsilon_k^{\max}$ (light bars) for pure
$k$-photon-subtracted states $\ket{\psi_k}$.} \label{fig:plhysto}
\end{figure}

We have shown that, quite remarkably, the simple measurement of the
CM of a pure PSS enables to pin down the entropy of entanglement
quantitatively within analytical {\em a priori} bounds. In fact, one
can appreciate how close the lower and upper bounds (both functions
of the second moments only) are  to each other for various values of
$k$ in Fig.~\ref{fig:bounds}[{\it (a)--(c)}]. A crucial fact is that
the absolute error $\Delta_k=(E^{\rm up}_k-E^{\rm low}_k)/2$ on the
entanglement quantification asymptotically saturates  (for $r
\rightarrow \infty$) to a constant value $$\Delta_k^{\max} =
\frac12[\log(4+8k) -1],$$ being $\Delta_k \le \Delta_k^{\max}$ for any
finite $r$. Accordingly, since the actual value of the entanglement
(measured by the average between lower and upper bound) diverges
linearly with the squeezing $r$, the relative error $\delta_k =
(E^{\rm up}_k-E^{\rm low}_k)/(E^{\rm up}_k+E^{\rm low}_k)$ on the
estimate of $E(\psi_k)$ from the CM {\em vanishes} for $r \gg 0$,
rendering our method rigorously accurate. We notice that, in
general, the error $\Delta_k$ increases with $k$, although it stays
of the order of few units -- on a scale ranging to infinity -- even
for big $k$ (e.g. $\Delta_k \le 4$ for up to $k=1000$ photon
subtractions per beam), thus scarcely affecting the quality of the
estimate [see Fig.~\ref{fig:bounds}]. We believe that a physical
explanation for the scaling of $\Delta_k$ is rooted in the fact that
with increasing $k$ the PSS $\ket{\psi_k}$ are increasingly more
non-Gaussian, hence there is more information not retrievable from
second moments only. This argument can be made quantitative by
evaluating the entropic non-Gaussianity \cite{parisentro}
$\Upsilon_k$  of $\ket{\psi_k}$, which simply amounts in this case
to the Von Neumann entropy of the associated Gaussian state with CM
$\sig_k$. We obtain, for $r \gg 0$, $$\Upsilon_k^{\max} = \frac{1}{2}
\left[\log \left(\frac{k}{2}\right)+\sqrt{2 k+1} \log
\left(\frac{k+\sqrt{2
   k+1}+1}{k}\right)\right].$$ For any $r$, the non-Gaussianity and
   the absolute error are very close to each other, with
   $\Upsilon_k \ge \Delta_k$, and exhibit the same scaling with $k$
   (see Fig.~\ref{fig:plhysto}). This fascinating connection adds
   insight to our analysis and leads us to sum up the results
   achieved so far as follows.
{\em Entanglement in ideal photon-subtracted states can be measured
from the covariance matrix up to a narrow  error that scales with
the states' non-Gaussianity}.

\section{Generalization to mixed states and further
remarks} In most practical implementations, efficient
photon-number-resolving detectors are not available and the
conditional generation of PSS is achieved by means of ``on/off''
type detectors which are only able to discriminate the vacuum from a
bunch of an undefined number of photons \cite{kitagawa}. This means
that a more appropriate description of a class of PSS must be in terms of statistical
mixtures of the form $\varrho_{\bar k} = \sum_k p_k
\ket{\psi_k}\!\bra{\psi_k}$, where we can define ${\bar k} \equiv
\sum_k p_k k$ as the `average' number of photons subtracted per beam
\cite{notesym}. For these mixed states not even the logarithmic
negativity is available in closed form (and its numerical evaluation
on a computer requires several days for any given squeezing degree
\cite{kitagawa}), let alone the entanglement of formation.
Remarkably, our bounds can be immediately extended to pinpoint the
 entanglement of formation of mixed symmetric PSS from the sole
knowledge of second moments. We first observe that (having zero
first moments) the CM transforms linearly: $\sig_{\bar k} = \sum_k
p_k \sig_k$, hence it can be computed analytically for any
probability distribution $\{p_k\}$ from
Eqs.~(\ref{eq:cx1},\ref{eq:cx3}). The $E_F$ of the corresponding
Gaussian state with CM $\sig_{\bar k}$, according to the extremality
theorem \cite{extra}, stands as a lower bound for the $E_F$ of the
non-Gaussian mixed PSS $\varrho_{\bar k}$ \cite{noteconj}. On the
other hand, denoting by $\gamma^{(\bar k)}$ the $(1,3)$ element of
the mixed-state CM $\sig_{\bar k}$, $\gamma^{(\bar k)} = \sum_k p_k
\gamma^{(k)}$, the upper bound \eq{eq:eup} is immediately extended
to the mixed case: $\log[1+2 \gamma^{(\bar k)}] \ge
E_F(\varrho_{\bar k})$. The proof follows from the concavity of the
log function, the convexity of the entanglement of formation, and
obviously Theorem 1. Namely, $\log[1+2 \gamma^{(\bar k)}] \ge \sum_k
p_k \log[1+2 \gamma^{(k)}] \ge \sum_k p_k E(\psi_k) \ge E_F
(\varrho_{\bar k})$. The behavior of the bounds in such more realistic conditions is shown in Fig.~\ref{fig:bounds}{\it(d)} for an instance
with ${\bar k}=1$. We observe, in general, that for reasonable
modeling of the mixture (e.g., $p_k$ following a binomial
distribution) the loosening of the bounds compared to the ideal
cases with $k=\lfloor{\bar k}\rfloor$ is negligible: our scheme is
efficient and {\em robust} against the specific source of imperfection considered here (the usage of non-phon-number-resolving detectors). We plan to deepen our investigation in the future following the experimental progresses in the generation of (generally non-symmetric) PSS states \cite{notesym}, thus properly modeling other sources of imperfections (e.g. mismatches or dark counts in the photon conditioning) that arise in practical demonstrations \cite{science2006}, in order to test the robustness and reliability of our techniques for estimating entanglement in fully realistic situations.

In this context, let us briefly comment on the direct implementation of our results in experiments. Once a
two-mode PSS is prepared, one needs to measure the full CM by
homodyne detections as  in \cite{francesi,porzio}, transform it in
standard form (i.e. extract the symplectic invariants
$a_1,a_2,\gamma_x,\gamma_p$) \cite{duan,ourreview}, and then (upon
verification that the standard-form CM has the structure predicted
here: a benchmark for the state engineering) readily evaluate our
lower and upper bounds to ensure an accurate estimate of the
entanglement of formation of the produced non-Gaussian state.

\section{Conclusion}
In this paper, in the spirit of \cite{extra,carles}, we have gone
beyond the conventional belief that out of Gaussian states the
covariance matrix plays a marginal role in CV entanglement
quantification. On the contrary, we demonstrated  that clever
exploitation of  such an easily accessible component of the state,
bears extremely useful and precise information on the quantification
of non-Gaussian entanglement produced in experiments, specifically
for the important class of (realistic) photon-subtracted states
\cite{photsub,kitagawa}. This is the start of a program which will
continue with the systematic investigation of quantitative
entanglement witnesses for other classes of non-Gaussian states in
terms of low-order moments. We hope with our result to stimulate
advances in the engineering and characterization of non-Gaussian
resources, and their exploitation for demonstrations impossible to
achieve with Gaussian states only, in order to explore the actual
limits that quantum mechanics poses on the access and manipulation
of information.

\acknowledgments{I thank C. Rod\'o, A. Sanpera, F. Dell'Anno, F.
Illuminati, V. D'Auria, V. De Angelis, P. Kosinar, P. Paule for
discussions.}


\begin{thebibliography}{99}
\bibitem{notelog}
Logarithms are taken in natural basis throughout the paper.
\bibitem{notesym}
Symmetry between the parties is typically required by quantum
protocols and should be fulfilled by experimental realizations, hence  we do
not consider here non-symmetric states where a different number of
photons is subtracted from each beam. However, since these instances can more closely resemble realistic implementations, we commit ourselves to extend our methods to encompass such a more general framework in further work.
\bibitem{noteconj}
The rigorous validity of this lower bound relies on the wide\-spread
conjecture on the additivity of $E_F$, which holds true for Gaussian
states \cite{efsym,marian}, but no proof is known for PSS.
\end{thebibliography}
\end{document}